\documentclass[12pt]{iopart}
\usepackage{graphicx}
\usepackage{iopams}

\begin{document}

\title[]{Parametrized post-Newtonian virial theorem}

\author{Mahmood Roshan}

\address{School of Physics, Institute for Research in Fundamental Sciences (IPM), P. O. Box 19395-5531, Tehran, Iran}
\ead{rowshan@alumni.ut.ac.ir}
\begin{abstract}
Using the parametrized post-Newtonian equations of hydrodynamics, we derive the tensor form of the parametrized post-Newtonian virial theorem.

\end{abstract}

 %Keywords required only for MST, PB, PMB, PM, JOA, JOB?
\submitto{\CQG}
\maketitle

\section{Introduction}

The virial theorem has a wide range of applications in physics. For example, the virial theorem is a powerful tool for studying static, non-evolving (relaxed) systems such as stars, gas clouds, star clusters, galaxies and galaxy clusters. It can also be used to study large structures in the universe. Not only is it applicable to nonrelativistic dynamical systems, but it can also be formulated to deal with relativistic systems which require special relativity or even general relativity for their description. Of course, one cannot expect to obtain a complete description of the system only by using the virial theorem. However, this theorem can provide considerable insight into the dynamical behavior of the system. For example, Zwicky could discover the large mass discrepancy in the galaxy clusters only by applying the virial theorem to the Coma cluster \cite{zwicky}.

For a historical review of the virial theorem and also its applications in stellar astrophysics, see \cite{collins}.

In the context of the alternative metric theories of gravity, it is known that only the values of a set of numerical coefficients in the components of the post-Newtonian metric vary from theory to theory. Thus, one can encompass a large class of alternative theories by introducing arbitrary parameters in place of the numerical coefficients. In the other words, this framework provides a model independent tool for testing the metric theories of gravity in the solar system in the small velocity and weak-field regime. Although this idea dates back to Eddington \cite{eddi}, but the parametrized post-Newtonian (PPN) formalism was developed by Nordtvedt and by Will (see \cite{willbook} for more details).

 In this paper, we derive the PPN virial theorem. In fact, our main result is a post-Newtonian virial theorem which can be used in a large class of metric theories of gravity. It should be noted that the post-Newtonian virial theorem in general relativity has already been investigated by Chandrasekhar \cite{chandra1,chandra}.

The outline of this paper is as follows: In section \ref{section1}, we derive the PPN hydrodynamics equations for metric theories of gravity. In section \ref{section2}, we derive the tensor and scalar form of the PPN virial theorem from the PPN hydrodynamics equations.
\section{Hydrodynamics equations in the post-Newtonian approximation}
\label{section1}

Einstein equivalence principle is the foundation of metric theories of gravity. Thus in these theories, the world lines of test bodies are geodesics of the metric $g_{\mu\nu}$. Mathematically, this postulate can be written as the following identity
\begin{eqnarray}
\nabla_{\mu}T^{\mu\nu}=0
\label{iden}
\end{eqnarray}
where $T^{\mu\nu}$ is the energy-momentum tensor. We assume that for a hydrodynamic system the energy-momentum tensor is given by
\begin{eqnarray}
T_{\mu\nu}=(\rho(1+\Pi)+p)u_{\mu}u_{\nu}+p g_{\mu\nu}
\end{eqnarray}
where $\rho$ is the rest-mass-energy density of atoms in the fluid element, $p$ is the pressure, $\Pi$ is the specific density of internal kinetic and thermal energy in the fluid element and $u^{\mu}$ is the four velocity of the fluid element in the comoving frame (we use the geometrized units in which the vacuum speed of light and the gravitational constant are unity i.e. $c=1$, $G=1$). In the post-Newtonian limit, $v^{2}$ ($\textbf{v}=\frac{d\textbf{x}}{dt}$), $\frac{p}{\rho}$, $\Pi$ and $U$ (Newtonian gravitational potential) are small quantities:
\begin{eqnarray}
v^2 \sim U \sim \Pi \sim \frac{p}{\rho} \sim O(2)
\end{eqnarray}
where $O(n)$ stands for a term of order $\frac{1}{c^n}$, where $c$ is the vacuum speed of light. On the other hand the PPN metric is given by \cite{willbook},
\begin{eqnarray}
\fl g_{00}=-1+2U-2\beta U^2-2\xi \Phi_{W}+(2\gamma+2+\alpha_3+\zeta_1-2\xi)\Phi_1  \label{ppnmetric1}& \\ \nonumber +2(3\gamma-2\beta+1+\zeta_2+\xi)\Phi_2 +2(1+\zeta_3)\Phi_3&\\+ 2(3\gamma+3\zeta_4-2\xi)\Phi_4-(\zeta_1-2\xi)\mathcal{A}\nonumber
\end{eqnarray}
\begin{eqnarray}
 \fl g_{0j}=-\frac{1}{2}(4\gamma+3+\alpha_1-\alpha_2+\zeta_1-2\xi)V_j-\frac{1}{2}(1+\alpha_2-\zeta_1+2\xi)W_j
 \label{ppnmetric2}
 \end{eqnarray}
 \begin{eqnarray}
 \fl g_{ij}=(1+2\gamma U)\delta_{ij}
 \label{ppnmetric3}
\end{eqnarray}
where $U$ is the Newtonian gravitational potential determined in terms of $\rho$ as
\begin{equation}
 U=\int \frac{\rho'}{|\textbf{x}-\textbf{x}'|}d^3x'
 \end{equation}
 and $\Phi_W$, $\Phi_1$, $\Phi_2$, $\Phi_3$, $\Phi_4$, $V_j$, $W_j$, and $\mathcal{A}$ are other metric potentials (see \ref{ap1} for the definition of these potentials). $\alpha_1$, $\alpha_2$, $\alpha_3$, $\zeta_1$, $\zeta_2$, $\zeta_3$, $\zeta_4$, $\xi$, $\gamma$ and $\beta$ are the PPN parameters. As it is known, each parameter has a special meaning and one can compare different metric theories in the post-Newtonian approximation only by comparing their PPN parameters.

Energy-momentum tensor $T_{\mu\nu}$ and the Christoffel symbols $\Gamma^{\lambda}_{\mu\nu}$ can be written to the post-Newtonian order by using the PPN metric. See table 4.1 of \cite{willbook} for $T_{\mu\nu}$ in the post-Newtonian limit and table 6.1 for Christoffel symbols. Now, by substituting the metric components (\ref{ppnmetric1})-(\ref{ppnmetric3}) into equation (\ref{iden}) and by using the tables 4.1 and 6.1 of \cite{willbook}, the time component of (\ref{iden}) takes the following form
\begin{equation}
\frac{\partial\sigma}{\partial t}+\frac{\partial (\sigma v^i)}{\partial x^i}+(3\gamma-2)\rho \frac{\partial U}{\partial t}-\frac{\partial p}{\partial t}+(3\gamma-3)\frac{\partial U}{\partial x^i}\rho v^{i}=0
\label{continuty}
\end{equation}
where
\begin{equation}
\sigma=\rho (1+\Pi+v^2+2U+\frac{p}{\rho})
\label{sssigma}
\end{equation}
and $\textbf{v}=\frac{d\textbf{x}}{dt}$ is the velocity of a fluid element. Equation (\ref{continuty}) is the continuity equation in the post-Newtonian approximation. It is straightforward to show that the space component of (\ref{iden}) takes the form
\begin{eqnarray}
\fl \frac{\partial(\sigma v_i)}{\partial t}+\frac{\partial}{\partial x^j}(\sigma v_i v_j)+(5\gamma-1)\rho v_i \left[\frac{\partial U}{\partial t}+v^j \frac{\partial U}{\partial x^j}\right]  \label{space} & \\+\rho\frac{\partial}{\partial x^i}\left[(\beta+\gamma)U^2+\xi \Phi_W-\Phi+\frac{1}{2}(\zeta_1-2\xi)\mathcal{A}\right] \nonumber & \\+(1-2\gamma U)\frac{\partial p}{\partial x^i}-(c_1+c_2)\rho v^j \left(\frac{\partial V_i}{\partial x^j}+ \frac{\partial V_j}{\partial x^i}\right)\nonumber  & \\
-(\sigma+\rho \gamma v^2)\frac{\partial U}{\partial x^i}-\frac{\partial}{\partial t}[c_1 W_i+c_2 V_i]=0\nonumber
\end{eqnarray}
 where
\begin{equation}
c_2=\frac{1}{2}(4\gamma+4+\alpha_1-\alpha_2+\zeta_1-2\xi)
\end{equation}
\begin{equation}
c_1=\frac{1}{2}(1+\alpha_2-\zeta_1+2\xi)
\end{equation}
and
\begin{eqnarray}
\fl \Phi=\frac{1}{2}(2\gamma+2+\alpha_3+\zeta_1-2\xi)\Phi_1+(1+\zeta_3)\Phi_3 \nonumber & \\+(3\gamma-2\beta+1+\zeta_2+\xi)\Phi_2+(3\gamma+3\zeta_4-2\xi)\Phi_4
\end{eqnarray}
In order to simplify equation (\ref{space}) let us introduce the potential $\chi$ as (\cite{willbook})
\begin{eqnarray}
\chi=-\int \rho(\textbf{x}')|\textbf{x}-\textbf{x}'|d^3x'
\label{komaki2}
\end{eqnarray}
one can easily verify that
\begin{eqnarray}
\frac{\partial^2\chi}{\partial t \partial x^j}=V_j-W_j
\label{komaki3}
\end{eqnarray}
on the other hand, for a perfect nonviscous fluid the Euler's equation in the Newtonian limit is given by
\begin{eqnarray}
 \rho \frac{d v_i}{dt}=\rho\frac{\partial U}{\partial x^i}-\frac{\partial p}{\partial x^i}+O(4)
 \label{komaki1}
 \end{eqnarray}
where $O(4)$ is the post-Newtonian corrections to the Newtonian Euler's equation. Note that we do not need the exact form of the post-Newtonian corrections $O(4)$.
Now, by using equations (\ref{komaki2})-(\ref{komaki1}), equation (\ref{space}) takes the following form
\begin{eqnarray}
\fl \frac{\partial(\sigma v_i)}{\partial t}+\frac{\partial}{\partial x^j}(\sigma v_i v_j)+\frac{\partial}{\partial x^i}[(1+3(\gamma-1)U)p]-p\frac{\partial U}{\partial x^i} \nonumber & \\
+\rho\frac{\partial}{\partial x^i}\left[\frac{1}{2}\zeta_1\mathcal{A}-\frac{\alpha_3-\zeta_1}{2}\Phi_1-\zeta_2\Phi_2-\zeta_3\Phi_3-3\zeta_4\Phi_4 +\frac{1}{2}(1+\alpha_2-\zeta_1)\chi_{,00}\right]
\nonumber & \\+\frac{4\gamma+4+\alpha_1}{2}\rho v^j \frac{\partial V_j}{\partial x^i}+\xi\rho\frac{\partial}{\partial x^i}\left[\chi_{,00}+\Phi_W-\mathcal{A}+\Phi_1-\Phi_2+2\Phi_4\right]\label{simspace} &\\
+(5\gamma-1)\rho\frac{d}{dt}\left[v_i U-\frac{4\gamma+4+\alpha_1}{2(5\gamma-1)}V_i\right]-2\rho\left[\hat{\phi}\frac{\partial U}{\partial x^i}+\frac{1}{2}\frac{\partial \hat{\Phi}}{\partial x^i}\right]=0\nonumber
\end{eqnarray}
where
\begin{equation}
\hat{\phi}=\frac{1+\gamma}{2}v^2+\frac{\Pi}{2}+\frac{3\gamma-2\beta+1}{2} U+\frac{3\gamma}{2}\frac{p}{\rho}
\label{hatphi1}
\end{equation}
\begin{equation}
\nabla^2\hat{\Phi}=-8\pi\rho\hat{\phi}
\label{hatphi2}
\end{equation}
Equation (\ref{simspace}) is the Eulerian equation in the post-Newtonian approximation. Equations (\ref{simspace}) and (\ref{continuty}) are enough for describing a hydrodynamic system in the post-Newtonian approximation.

\section{PPN virial theorem}
\label{section2}
In Newton's theory of gravity, the total energy of a gravitating system is conserved. Furthermore, if the gravitating system is not affected by external forces, its total linear and angular momentum are also conserved. However, in the post-Newtonian approximation this is not the case for all metric theories of gravity \cite{willbook}. For example, for the conservation of the total linear momentum of system, the parameters $\zeta_1,\zeta_2,\zeta_3,\zeta_4$ and $\alpha_3$ should be zero \cite{willbook}. It is obvious that, for theories which violate the conservation of the total linear momentum, we can not find a PPN virial theorem. Because in these theories, the given gravitating system would not be stationary or periodic in time in the post-Newtonian approximation and consequently it can not be considered as a virialized system. Thus we shall restrict ourselves to metric theories in which five PPN parameters $\zeta_1,\zeta_2,\zeta_3,\zeta_4$ and $\alpha_3$ are zero.

Now, let us rewrite equation (\ref{simspace}) as follows
 \begin{eqnarray}
\fl \frac{\partial(\sigma v_i)}{\partial t}+\frac{\partial}{\partial x^j}(\sigma v_i v_j)+\frac{\partial}{\partial x^i}[(1+3(\gamma-1)U)p]-p\frac{\partial U}{\partial x^i}  \label{sim2space} & \\ +\frac{1+\alpha_2}{2}\rho\left[\frac{d}{dt}(V_i-W_i)-\mathcal{W}_{i}\right]+\frac{4\gamma+4+\alpha_1}{2}\rho v^j \frac{\partial V_j}{\partial x^i}\nonumber &\\+(5\gamma-1)\rho\frac{d}{dt}\left[v_i U-\frac{4\gamma+4+\alpha_1}{2(5\gamma-1)}V_i\right]\nonumber & \\
+\xi\rho\frac{\partial}{\partial x^i}[\mathcal{B}+\Phi_W-\mathcal{A}-\Phi_2+2\Phi_4]=0\nonumber
\end{eqnarray}
 where we have used the following expressions
 \begin{equation}
\mathcal{B}=\chi_{,00}-\mathcal{A}+\Phi_1
\end{equation}
\begin{equation}
 \mathcal{W}_i=v^j \frac{\partial}{\partial x^j}(V_i-W_i)
 \end{equation}
In order to find the PPN virial theorem, we multiply equation (\ref{sim2space}) by $x_j$ and integrate over the volume $V$ of the fluid. Most integrals can be
 calculated by using the lemmas introduced in \cite{chandra}. So we do not present the detail of the calculations here. However, there are four new integrals which we combine them in the following expression
 \begin{equation}
 M_{ij}=-\int x_j\rho\frac{\partial}{\partial x_i}[\mathcal{B}+\Phi_{W}-\Phi_2+2\Phi_4]d^3 x
 \label{m}
 \end{equation}
in the \ref{ap1} we have shown that $M_{ij}$ is a symmetric tensor. Hence, the result can be written as
 \begin{eqnarray}
 \fl \frac{d}{dt}\int x_j \Pi_i d^3x=2\mathcal{T}_{ij}+\mathfrak{M}_{ij}+\delta_{ij}P-\frac{4\gamma+3+\alpha_1-\alpha_2}{2} (\mathfrak{u}_{ij}+\mathfrak{U}_{ij}) \label{vir1} &\\ +2\hat{\Phi}_{ij}+(5\gamma-1)W_{ij}-\frac{1+\alpha_2}{2}Q_{ij} +\frac{3(1+\alpha_2)}{4}Z_{ij}+\xi M_{ij} \nonumber
 \end{eqnarray}
 where
\begin{equation}
\Pi_i=\sigma v_i+(5\gamma-1)\rho\left(v_i U-\frac{4\gamma+4+\alpha_1}{2(5\gamma-1)}V_i\right)+\frac{1+\alpha_2}{2}\rho(V_i-W_i)
 \end{equation}
 and $\int \Pi_i d^3x$ is the conserved total linear momentum of the system \cite{willbook}. In fact, by integrating equation (\ref{sim2space}) over the volume
 of the fluid, one can verify that $\frac{d}{dt}\int \Pi_i d^3x=0$. The tensors on the right-hand side of equation (\ref{vir1}) are defined as follows
\begin{eqnarray}
\fl \mathcal{T}_{ij}=\frac{1}{2}\int\sigma v_i v_j  d^3x
&\\
\fl \delta_{ij}P=\delta_{ij}\int [(1+(3\gamma-1)U)p]d^3x
&\\
\fl \mathfrak{M}_{ij}=\int x_j\rho \frac{\partial U}{\partial x^i}d^3x
&\\
\fl \mathfrak{U}_{ij}=\int \rho x_j v^k \frac{\partial V_k}{\partial x^i}d^3x
&\\
\fl \hat{\Phi}_{ij}=\int x_j \rho \left[\hat{\phi}\frac{\partial U}{\partial x^i}+\frac{1}{2}\frac{\partial \hat{\Phi}}{\partial x^i}\right]d^3x
&\\
\fl W_{ij}=\int \rho U v_i v_j d^3x
&\\
 \fl \mathfrak{u}_{ij}=\int \rho v_j V_i d^3x
&\\
 \fl Q_{ij}=\int\int \rho(\textbf{x})\rho(\textbf{x}')\textbf{v}(\textbf{x}').(\textbf{x}-\textbf{x}')
\frac{v_i(\textbf{x})(x_j-x'_j)+v_j(\textbf{x})(x_i-x'_i)}{|\textbf{x}-\textbf{x}'|^3}d^3xd^3x'
&\\
\fl Z_{ij}=\int\int \rho(\textbf{x})\rho(\textbf{x}')\textbf{v}(\textbf{x}).(\textbf{x}-\textbf{x}')\textbf{v}(\textbf{x}').(\textbf{x}-\textbf{x}')
\frac{(x_i-x'_i)(x_j-x'_j)}{|\textbf{x}-\textbf{x}'|^5}d^3xd^3x'
& \\
\fl M_{ij}=-\int\int\int\rho(\textbf{x})\rho(\textbf{x}')\rho(\textbf{x}'')[(\frac{3(\textbf{x}-\textbf{x}').(\textbf{x}-\textbf{x}'')}{|\textbf{x}'-\textbf{x}''|
|\textbf{x}'-\textbf{x}|^5}\nonumber & \\+\frac{(\textbf{x}-\textbf{x}'').(\textbf{x}'-\textbf{x}'')}{|\textbf{x}-\textbf{x}'|^3|\textbf{x}'-\textbf{x}''|})
(x_i-x'_i)(x_j-x'_j) \nonumber & \\+\frac{(x_i-x''_i)(x_j-x'_j)+(x_j-x''_j)(x_i-x'_i)}{|\textbf{x}-\textbf{x}'|^3|\textbf{x}'-\textbf{x}''|}]d^3xd^3x'd^3x''
 \end{eqnarray}
 These tensors are symmetric in $i$ and $j$ (see \cite{chandra} and \ref{ap1} for more details). Since the right-hand side of equation (\ref{vir1}) is symmetric then the antisymmetric part of the left-hand side of (\ref{vir1}) should vanish, i.e.
 \begin{equation}
 \frac{d}{dt}\int (x_j\Pi_i-x_i\Pi_j)d^3x=0
 \end{equation}
 this equation expresses the conservation of the total angular momentum of the system. If the system is stationary or periodic in time then the time average over a long time interval of the symmetric part of the left-hand side of equation (\ref{vir1}) vanishes, i.e.
 \begin{equation}
 \left\langle\frac{d}{dt}\int (x_j\Pi_i+x_i\Pi_j)d^3x\right\rangle\simeq0
 \end{equation}
where $\langle...\rangle$ denotes the time average. Thus the PPN generalization of the classical tensor virial theorem takes the following form
 \begin{eqnarray}
\fl 2\left\langle\mathcal{T}_{ij}\right\rangle+\left\langle\mathfrak{M}_{ij}\right\rangle+\delta_{ij}\left\langle P\right\rangle-\frac{4\gamma+3+\alpha_1-\alpha_2}{2}\left\langle\mathfrak{u}_{ij}+\mathfrak{U}_{ij}\right\rangle
 +2\left\langle\hat{\Phi}_{ij}\right\rangle+(5\gamma-1)\left\langle W_{ij}\right\rangle \label{tensorvir} & \\-\frac{1+\alpha_2}{2}\left\langle Q_{ij}\right\rangle +\frac{3(1+\alpha_2)}{4}\left\langle Z_{ij}\right\rangle+\xi \left\langle M_{ij}\right\rangle=0\nonumber
 \end{eqnarray}
 By contracting $i$ and $j$, the scalar PPN virial theorem is
 \begin{eqnarray}
 \fl 2\left\langle\mathcal{T}\right\rangle+\left\langle\mathfrak{M}\right\rangle+3\left\langle P\right\rangle-\frac{4\gamma+3+\alpha_1-\alpha_2}{4}\left\langle\mathfrak{u}\right\rangle\nonumber & \\
 +2\left\langle\hat{\Phi}\right\rangle+(5\gamma-1)\left\langle W\right\rangle-\frac{1+\alpha_2}{4}\left\langle Z\right\rangle+\xi \left\langle M\right\rangle=0
 \label{scalarvir}
 \end{eqnarray}
 where
 \begin{eqnarray}
\fl \left\langle\mathcal{T}\right\rangle =\frac{1}{2}\left\langle\int \sigma v^2 d^3x\right\rangle &\\ \fl \left\langle\mathfrak{M}\right\rangle=-\frac{1}{2}\left
\langle\int \rho U d^3x\right\rangle &\\ \fl \left\langle W \right\rangle=\left\langle\int \rho v^2 U d^3x\right\rangle &\\ \fl \left\langle\mathfrak{u}\right\rangle=-2\left\langle\mathfrak{U}\right\rangle=\left\langle\int\rho v_i V_i d^3x\right\rangle &\\ \fl
\left\langle\hat{\Phi}\right\rangle=-\left\langle\rho\hat{\phi}U d^3x\right\rangle &\\ \fl
\left\langle Q\right\rangle =2\left\langle Z\right\rangle=2\left\langle\int\int \rho(\textbf{x})\rho(\textbf{x}')\frac{\textbf{v}(\textbf{x}).(\textbf{x}-\textbf{x}')\textbf{v}(\textbf{x}').
(\textbf{x}-\textbf{x}')}{|\textbf{x}-\textbf{x}'|^3}d^3xd^3x'\right\rangle &\\ \fl \left\langle M\right\rangle =-2\left\langle\int\int\int \rho(\textbf{x})\rho(\textbf{x}')\rho(\textbf{x}'') \frac{(\textbf{x}-\textbf{x}'').(\textbf{x}'-\textbf{x}'')}{|\textbf{x}-\textbf{x}'|^3|\textbf{x}'-\textbf{x}''|}d^3xd^3x'd^3x''\right\rangle
 \label{vir2}
 \end{eqnarray}
Equations (\ref{tensorvir}) and (\ref{scalarvir}) are our final results. It is clear from equation (\ref{scalarvir}) that the scalar PPN virial theorem has six extra terms relative to the Newtonian virial theorem. If we set $\gamma=\beta=1$ and $\alpha_1=\alpha_2=\xi=0$, then we obtain the post-Newtonian virial theorem in general relativity. In this case, our result is identical to the result obtained by Chandrasekhar \cite{chandra}.

It is obvious that the PPN virial theorem can be applied to investigate the post-Newtonian effects of metric theories of gravity on the hydrodynamic behavior of the systems which are in the post-Newtonian limit. By applying the PPN virial theorem to such cases, one can compare metric theories of gravity with general relativity. In these systems, general relativistic effects on the dynamics cannot be ignored. For example, post-Newtonian corrections in supermassive stars and massive white dwarfs would lead to unexpected gravitational instabilities. This fact has been shown by means of rather detailed calculations \cite{fowler,tooper}. However, by using the post-Newtonian virial theorem this result could be anticipated without the need of detailed calculations \cite{collins,fricke}. Of course, as we mentioned before, by using the virial theorem one cannot expect to obtain as complete a description of a hydrodynamic system as would be possible from the solution of the field equations. However, the virial theorem can provide extensive insight into the behavior of the system.

\section{Conclusion}
 In this paper, we have derived the PPN virial theorem. The virial theorem in the post-Newtonian approximation has already been derived only in general relativity. We have generalized it to every metric theory of gravity. Albeit, we mentioned that the PPN virial theorem can only be written for theories in which parameters $\zeta_1$, $\zeta_2$, $\zeta_3$, $\zeta_4$ and $\alpha_3$ are zero.

\appendix
\section{Tensor $M_{ij}$}
\label{ap1}
In this appendix we simplify $M_{ij}$ (integral (\ref{m})) as far as possible. Also we show that $M_{ij}$ is a symmetric tensor. First, let us write the metric
potentials of the PPN metric (see \cite{willbook} for more details)
\begin{eqnarray}
\Phi_W=\int \rho'\rho''\frac{\textbf{x}-\textbf{x}'}{|\textbf{x}-\textbf{x}'|^3}.\left(\frac{\textbf{x}'-\textbf{x}''}{|\textbf{x}-\textbf{x}''|}-\frac{\textbf{x}-
\textbf{x}''}{|\textbf{x}'-\textbf{x}''|}\right)d^3xd^3x'&\\
\Phi_1=\int \frac{\rho' v'^2}{|\textbf{x}-\textbf{x}'|}d^3x'&\\
\Phi_2=\int \frac{\rho' U'}{|\textbf{x}-\textbf{x}'|}d^3x'& \\
\Phi_3=\int \frac{\rho' \Pi'}{|\textbf{x}-\textbf{x}'|}d^3x'&\\
\Phi_4=\int \frac{p'}{|\textbf{x}-\textbf{x}'|}d^3x'&\\
\mathcal{A}=\int \frac{\rho'[\textbf{v}'.(\textbf{x}-\textbf{x}')]^2}{|\textbf{x}-\textbf{x}'|^3}d^3x'&\\ \mathcal{B}=\int \frac{\rho'(\textbf{x}-\textbf{x}')}
{|\textbf{x}-\textbf{x}'|}.\frac{d\textbf{v}'}{dt}d^3x'&\\
V_j=\int \frac{\rho' v'_j}{|\textbf{x}-\textbf{x}'|}d^3x'&\\ W_j=\int \frac{\rho' \textbf{v}'.(\textbf{x}-\textbf{x}')(x_j-x'_j)}{|\textbf{x}-\textbf{x}'|^3} d^3x'
\end{eqnarray}
where prim means $f'=f(\textbf{x}')$.

In order to simplify equation (\ref{m}), we start with its first term. Using the definition of $\mathcal{B}$ and the Euler's equation (\ref{komaki1}) we get
\begin{eqnarray}
\fl \int x_j \rho \frac{\partial \mathcal{B}}{\partial x^i}d^3x=-\int \left[\frac{x_j\rho}{|\textbf{x}-\textbf{x}'|}\frac{\partial p}{\partial x'^i}
-\rho \frac{x_j (x_i-x'_i)}{|\textbf{x}-\textbf{x}'|^3}(x_k-x'_k)\frac{\partial p}{\partial x'^k}
\right]d^3xd^3x'\nonumber & \\
 +\int \rho\rho'\rho''\frac{x_j (x_i-x'_i)(\textbf{x}-\textbf{x}').(\textbf{x}'-\textbf{x}'')}{|\textbf{x}-\textbf{x}'||\textbf{x}'-\textbf{x}''|^3}
d^3xd^3x'd^3x''\nonumber &\\
-\int \rho\rho'\rho''\frac{x_j(x'_i-x''_i)}{|\textbf{x}-\textbf{x}'||\textbf{x}'-\textbf{x}''|^3}d^3xd^3x'd^3x''
\end{eqnarray}
using integration by parts, one can easily show that the first integral on the right-hand side can be written as
\begin{equation}
2\int\rho\left[\delta_{ij}\frac{p(\textbf{x}')}{|\textbf{x}-\textbf{x}'|}-\frac{\partial}{\partial x^i}\left(\frac{x_j p(\textbf{x}')}{|\textbf{x}-\textbf{x}'|}
\right)\right]d^3xd^3x'
\end{equation}
on the other hand, by using the definition of $\Phi_4$, the above equation is equal to
\begin{equation}
-2\int x_j \rho \frac{\partial \Phi_4}{\partial x^i} d^3x
\end{equation}
Thus we can write
\begin{eqnarray}
\fl \int x_j \rho \frac{\partial \mathcal{B}}{\partial x^i}d^3x +2\int x_j \rho \frac{\partial \Phi_4}{\partial x^i}=\nonumber &\\  +\int \rho\rho'\rho''\frac{x_j (x_i-x'_i)(\textbf{x}-\textbf{x}').(\textbf{x}'-\textbf{x}'')}{|\textbf{x}-\textbf{x}'||\textbf{x}'-\textbf{x}''|^3}
d^3xd^3x'd^3x''\label{B} &\\
-\int \rho\rho'\rho''\frac{x_j(x'_i-x''_i)}{|\textbf{x}-\textbf{x}'||\textbf{x}'-\textbf{x}''|^3}d^3xd^3x'd^3x''\nonumber
\end{eqnarray}
Now consider the second term of equation (\ref{m}). Using the definition of $\Phi_W$ it is straightforward to show that
\begin{eqnarray}
\fl \int x_j \rho \frac{\partial \Phi_W}{\partial x^i}d^3x= \int \rho\rho'\rho'' x_j H_i(\textbf{x},\textbf{x}',\textbf{x}'')d^3xd^3x'd^3x''+\int
x_j\rho\frac{\partial \Phi_2}{\partial x^i}d^3x \label{W}&\\-
\int \rho\rho'\rho''  \frac{x_j(x_i-x''_i)(\textbf{x}-\textbf{x}').(\textbf{x}'-\textbf{x}'')}{|\textbf{x}-\textbf{x}'|^3|\textbf{x}-\textbf{x}''|^3}d^3xd^3x'd^3x''\nonumber
\end{eqnarray}
where $H_{i}(\textbf{x},\textbf{x}',\textbf{x}'')$ is given by
\begin{eqnarray}
\fl H_i=\frac{3(x_i-x'_i)(\textbf{x}-\textbf{x}')}{|\textbf{x}-\textbf{x}'|^5}.\left(\frac{(\textbf{x}-\textbf{x}'')}{|\textbf{x}'-\textbf{x}''|}-
\frac{(\textbf{x}'-\textbf{x}'')}{|\textbf{x}-\textbf{x}''|}\right) \nonumber &\\+\left(\frac{(x'_i-x''_i)}{|\textbf{x}-\textbf{x}'|^3|\textbf{x}-\textbf{x}''|}-\frac{(x_i-x''_i)}{|\textbf{x}-\textbf{x}'|^3|\textbf{x}'-\textbf{x}''|}\right)
\label{H}
\end{eqnarray}
it is obvious that $H_{i}(\textbf{x},\textbf{x}',\textbf{x}'')=-H_{i}(\textbf{x}',\textbf{x},\textbf{x}'')$. Now we can simplify the first integral on the right-hand side of (\ref{W}). Since the integrand is symmetric in $x$ and $x'$ (and also in $x''$ and $x'$) then the integral can be rewritten as
 \begin{eqnarray}
 \fl \int \rho\rho'\rho'' x_j H_i(\textbf{x},\textbf{x}',\textbf{x}'')d^3xd^3x'd^3x''= \nonumber & \\
 +\int \rho\rho'\rho'' \frac{3x'_j(x_i-x'_i)(\textbf{x}-\textbf{x}'').(\textbf{x}'-\textbf{x}'')}{|\textbf{x}-\textbf{x}'|^3|\textbf{x}'-\textbf{x}''|}d^3xd^3x'd^3x''\label{new2} &\\- \int \rho\rho'\rho'' \frac{(x_j-x'_j)(x_i-x''_i)}{|\textbf{x}-\textbf{x}'|^3|\textbf{x}'-\textbf{x}''|}d^3xd^3x'd^3x''\nonumber
\end{eqnarray}
 Similarly the third term on the right-hand side of (\ref{W}) can be written as
 \begin{eqnarray}
-\int \rho\rho'\rho''  \frac{x'_j(x_i-x'_i)(\textbf{x}-\textbf{x}'').(\textbf{x}'-\textbf{x}'')}{|\textbf{x}-\textbf{x}'|^3|\textbf{x}'-\textbf{x}''|^3}d^3xd^3x'd^3x''
\label{new1}
\end{eqnarray}
Also, the second integral on the right-hand side of (\ref{B}) can be replaced by
\begin{eqnarray}
\int \rho\rho'\rho''\frac{x''_j(x_i-x'_i)}{|\textbf{x}-\textbf{x}'|^3|\textbf{x}'-\textbf{x}''|}d^3xd^3x'd^3x''
\label{new3}
\end{eqnarray}
Now, by using equations (\ref{B})-(\ref{new3}) and (\ref{m}) we get
\begin{eqnarray}
\fl M_{ij} =
 -\int\int\int \rho\rho'\rho''\frac{3(x_i-x'_i)(x_j-x'_j)(\textbf{x}-\textbf{x}').(\textbf{x}-\textbf{x}'')}{|\textbf{x}'-\textbf{x}''||\textbf{x}-\textbf{x}'|^5}
d^3xd^3x'd^3x'' &\\ -\int\int\int \rho\rho'\rho''
\frac{(x_i-x'_i)(x_j-x'_j)(\textbf{x}-\textbf{x}'').(\textbf{x}'-\textbf{x}'')}{|\textbf{x}-\textbf{x}'|^3|\textbf{x}'-\textbf{x}''|^3}d^3xd^3x'd^3x''\nonumber &\\ +\int\int\int \rho\rho'\rho''\frac{(x_i-x''_i)(x_j-x'_j)+(x_j-x''_j)(x_i-x'_i)}{|\textbf{x}-\textbf{x}'|^3|\textbf{x}'-\textbf{x}''|}]d^3xd^3x'd^3x''\nonumber
\label{mm}
\end{eqnarray}
it is obvious from this equation that $M_{ij}$ is a symmetric tensor.

\end{document}